\documentclass[conference]{IEEEtran}

\IEEEoverridecommandlockouts
\usepackage{cite}
\usepackage{amsmath,amssymb,amsfonts}
\usepackage{algorithmic}
\usepackage{url}
\usepackage{graphicx}
\usepackage{textcomp}
\usepackage{pgfplots}
\usepackage{xcolor}
\begin{document}



\title{\texttt{minPIC}: Optimal Power Allocation in Multi-User Interference Channels}
\author{%
  \IEEEauthorblockN{Sagnik Bhattacharya\textsuperscript{*}, Abhiram Rao Gorle\textsuperscript{*}, John M. Cioffi}\footnote{}
  \IEEEauthorblockA{\textit{Department of Electrical Engineering}, \textit{Stanford University}\\
  Stanford, CA \\
  {sagnikb, abhiramg, cioffi}@stanford.edu}
  \IEEEauthorblockA{\small \textsuperscript{*}Equal contribution}
}

\maketitle

\begin{abstract}

 6G envisions massive cell-free networks with spatially nested multiple access (MAC) and broadcast (BC) channels without centralized coordination. This renders optimal resource allocation—across power, subcarriers, and decoding orders crucial for interference channels (ICs), where neither transmitters nor receivers can cooperate. Current orthogonal multiple access (OMA) methods, as well as non-orthogonal (NOMA) and rate-splitting (RSMA) schemes, rely on fixed heuristics for interference management, leading to suboptimal rates, power inefficiency, and scalability issues. This paper proposes a novel \texttt{minPIC} framework for optimal power, subcarrier, and decoding order allocation in general multi-user ICs. Unlike existing methods, \texttt{minPIC} eliminates heuristic SIC order assumptions. Despite the convexity of the IC capacity region, fixing an SIC order induces non-convexity in resource allocation, traditionally requiring heuristic approximations. We instead introduce a dual-variable-guided sorting criterion to identify globally optimal SIC orders, followed by convex optimization with auxiliary log-det constraints, efficiently solved via binary search. We also demonstrate that \texttt{minPIC} could potentially meet the stringent high‑rate, low‑power targets of immersive XR and other 6G applications. To the best of our knowledge, \texttt{minPIC} is the first algorithmic realisation of the Pareto boundary of the SIC-achievable rate region for Gaussian ICs, opening the door to scalable interference management in cell‑free networks.

\end{abstract}


\section{Introduction}
The emergence of next-generation wireless systems, especially for 6G, has sparked significant interest in addressing complex multi-user interference. Central to this is the multi-user interference channel (IC), a fundamental model where multiple transmitters independently send messages to their receivers at the same time, without coordination in encoding or modulation. Each transmitter splits its message into sub-user components, creating a complex web of transmitted signal interactions.



A multi-user IC models a network of $U$ transmitters and $U$ receivers communicating simultaneously without centralized coordination. Each transmitter $u$ splits its message into $U$ \textit{sub-user components}, resulting in $U^2$ transmitted signals. At receiver $i$, \textit{successive interference cancellation} (SIC) decodes its intended components and, based on decoding order, may also decode selected interfering components from other users.

In future 6G networks, cell-free massive MIMO—with many distributed access points instead of centralized base stations—will be key, making interference channel (IC) models highly relevant. Without central coordination, users must manage interference from multiple sources, requiring precise power and interference control—motivating problems like \texttt{minPIC}. These distributed, interference-aware architectures are driven by 6G metaverse demands (AR, VR, MR), which require per-user rates of hundreds of Mbps~\cite{3GPPTR38901,ITU-RM2410-0}, sub-10 ms latency~\cite{METISDeliverable17}, and 99.999\% reliability~\cite{ITURP53017}. Rising user density and limited spectrum~\cite{FCCSpectrumReport} make inter-user interference a major bottleneck, where traditional time/frequency division fails. Instead, smart power allocation across subcarriers or sub-user components enables efficient rate satisfaction in interference-heavy, cell-free MIMO environments.

\textbf{Related Work:
}The problem of characterizing the capacity region of the general Gaussian interference channel has been open for decades, with exact solutions known only in special cases (strong interference, deterministic) and two‐user channels to within one bit \cite{EtkinTseWang2008}. The Han–Kobayashi scheme remains the largest known achievable region, though intractable in general \cite{HanKobayashi1981}. Iterative water‐filling for parallel ICs \cite{ScutariPalomarBarbarossa2009} and branch‐and‐bound methods for small ICs \cite{JorswieckWiesel2008} yield local/global optima under fixed decoding. Recent works ~\cite{Clerckx2022, Clerckx2021RSMA_6G_ISAC} highlight RSMA’s (rate-splitting multiple access) versatility in 6G by enabling robust partial interference decoding across varied use-cases.

The novel \texttt{minPIC} (\underline{min}imum \underline{P}ower in \underline{I}nterference \underline{C}hannel) framework tackles key challenges by optimally allocating power across all $U^2$ sub-user components to minimize total transmit power while meeting per-user data-rate requirements. Unlike prior downlink-focused works with fixed splits, \texttt{minPIC} jointly optimizes uplink power allocation and SIC decoding order without preset splits or beamforming. At its core, the \texttt{minPIC} problem addresses two central challenges in next-gen wireless systems: allocating transmit power across sub-user components and sequencing SIC decoding to minimize total power--critical for meeting the demands of AR/VR and other data-intensive 6G applications.
\textbf{Our main contributions are twofold}: we establish the optimality of Gaussian signaling in the multi-user IC with SIC, forming a theoretical basis for algorithm design, and we propose an efficient algorithm to solve the \texttt{minPIC} problem and validate its performance in realistic interference-heavy scenarios.
The following sections present our algorithmic strategies, theoretical foundations, and numerical solutions for solving \texttt{minPIC} in various interference-limited environments.

\section{Problem Formulation}
\label{sec:prob-formulate}
We consider a $U$–user interference channel where each transmitter sends $U$ independently coded sub‐streams, totaling $U^2$ components. Let $x_{i,j}$ denote the $j$-th sub‐stream from user $i$, with transmit covariance $\mathbf{R}{xx}(i,j,n)\succeq\mathbf{0}$ on resource block $n$ $(n=0,\dots,N-1)$. Each receiver uses successive interference cancellation (SIC) to decode its own sub‐streams and may also decode selected interfering sub‐streams ${x{k,i}}_{k\neq i}$ to reduce residual interference.


Let $\Pi_u=(\pi_u(1),\dots,\pi_u(|D_u|))$ denote a decoding order at receiver $u$, and let $D_u\bigl(\Pi,p,b\bigr)\subseteq\{1,\dots,U\}$
be the set of users whose sub‐streams are decoded (and canceled) at receiver $u$ under order $\Pi_u$, power allocation $p$ and rate vector $b$.  After canceling all streams in $D_u$, receiver $u$ decodes its own sub‐streams $\{x_{u,j}\}_{j=1}^U$, achieving rates $b_{u,j}$ and aggregate rate 
$b_u \;=\;\sum_{j=1}^U b_{u,j}\;\ge\;b_{\min,u}$,
where $b_{\min,u}$ is the minimum rate requirement for user $u$.
The objective is to minimize the weighted sum-power while satisfying all users' rate constraints and ensuring covariance matrices remain positive semidefinite, under a choice of feasible SIC orders. This can formally be written as:
\begin{equation}\label{eq:opt}
\begin{aligned}
\min_{\{\mathbf{R}_{xx}(i,j,n)\succeq 0\}}
\;& \sum_{n=0}^{N-1}\sum_{u=1}^U w_u \,\mathrm{tr}\Bigl\{\sum_{j=1}^U\mathbf{R}_{xx}(u,j,n)\Bigr\}\\
\text{s.t.}\quad
& \text{C1: } b_u=\sum_{j=1}^U b_{u,j}\ge b_{\min,u},\text{ } \forall\,u=1,\dots,U,\\
& \text{C2: } \mathbf{R}_{xx}(u,j,n)\succeq 0,\text{ } \forall\,u,j,n,\\
& \text{C3: }\Pi_u\text{: feasible SIC order at RCVR }u,\text{ }\forall\,u.
\end{aligned}
\end{equation}

Here, $w_u>0$ is a weighting factor reflecting user priorities.  The first constraint enforces each user’s QoS rate, the second ensures valid transmit covariances, and the third captures the combinatorial SIC decoding order choices. Next, in Section III, we outline a proof sketch to show the optimality of Gaussian codebooks in achieving capacity for a 2-user Gaussian interference channel.

\section{Optimality of Gaussian Codebooks}
\label{sec:problem}

\begin{figure}[h]
    \centering
    \includegraphics[width=0.9\linewidth]{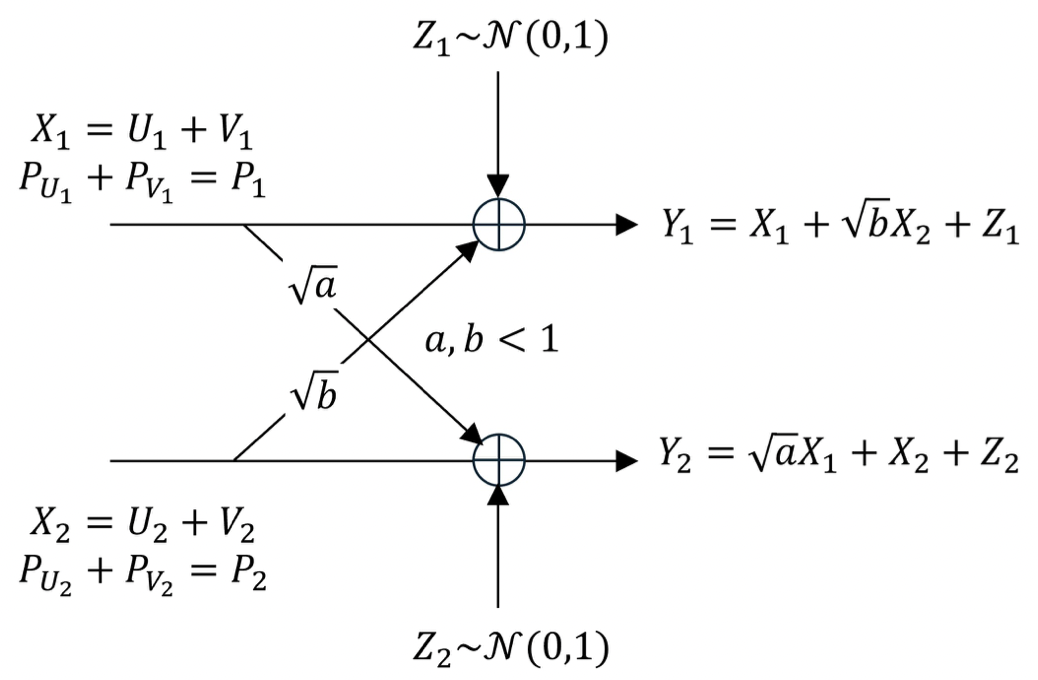}
    \caption{2-user Gaussian IC}
    \label{fig:2usergic}
\end{figure}

Figure \ref{fig:2usergic} considers the Gaussian Interference Channel (GIC) model from \cite{khandani2025}. In this proof sketch, we use a variational calculus approach to show that the capacity region of a 2-user weak GIC is achieved using Gaussian code-books.

\subsection{Setup}


Consider a two-users GIC with inputs \( X_1 \), \( X_2 \) and outputs \( Y_1 \), \( Y_2 \), defined as $Y_1 = X_1 + \sqrt{b}X_2 + Z_1, Y_2 = \sqrt{a}X_1 + X_2 + Z_2$, 
where \( a, b < 1 \) and \( Z_1, Z_2 \) are AWGN terms terms of zero mean and unit variance. $U_1, U_2, V_1, V_2$ denote the sub-user components in this case. The problem of finding the corresponding capacity region formulates as:
\begin{align}
    \text{Maximize:} \quad & R_1 + \mu R_2 = R_{U_1} + R_{V_1} + \mu (R_{U_2} + R_{V_2}) \notag \\
    \text{Subject to:} \quad & P_{U_1} + P_{V_1} = P_1, P_{U_2} + P_{V_2} = P_2
    \label{eq:optimization-problem}
\end{align}

Many interference-channel strategies, such as incremental methods for tracing the rate-region boundary, usually define a \textit{rate difference} or \textit{ratio} functional: $\Upsilon = \frac{N({f_i})}{D({f_i})},
$
where both $N(\cdot)$ and $D(\cdot)$ are linear combinations of integrals involving convolutions, such as $\int (f_1 * f_2 * n), \log(f_1 * f_2 * n), dx$ or $\int (f_{V_1} * f_{V_2} * \delta_{\text{noise}}), \log(\cdot), dx$, possibly with extra terms. These expressions capture sum variables like $V_1 + \sqrt{b}V_2 + Z$, where each user's codeword density is convolved with others and with noise. In later sections, minimizing such a functional turns out to reduce to finding the optimal $f$.

\subsection{Variational Derivative Under a Small Perturbation}
Let $f$ be the density of a sub-user component (e.g., $U_2$ or $V_1$), with all other codeword densities held fixed. We apply a small perturbation to $f$ by adding $\ell h$, yielding $\tilde{f}(\ell) = f + \ell h$, where $h$ is a real-valued function satisfying $\int h(x),dx = 0$ to preserve normalization. For sufficiently small $|\ell|$, $\tilde{f}(\ell) \ge 0$, ensuring it remains a valid probability density. This setup follows standard variational calculus methods \cite{Sagan1992ICV}, and a second-moment (power) constraint will be imposed separately.





\subsection{First-Order Expansion of Terms Involving $f$}

Consider an entropy term of the form  $  \int (f \ast \alpha)(x)\log(f \ast \alpha)(x)\,dx,$ where $\alpha$ denotes a fixed convolution of other densities. Define 
\[
  G(\ell) = \int ((f + \ell h) \ast \alpha)\, \log((f + \ell h) \ast \alpha)\, dx.
\]
Computation of $\frac{d}{d\ell}G(\ell)|_{\ell=0}$ uses a standard identity for integrals with $x\log x$ terms, like in \cite{agakov2004variational}, as follows:
\[
  \frac{d}{d\ell}(x\log x)\big|_{\ell=0} = \frac{dx}{d\ell} \log x \big|_{\ell=0} + \frac{dx}{d\ell} \big|_{\ell=0},
\]  
since $\frac{d}{dx}(x\log x) = \log x + 1$. Applying this gives: 
\begin{equation}
  \frac{d}{d\ell} G(\ell)\big|_{\ell=0} = \int (h \ast \alpha)(x)\, [\log(f \ast \alpha)(x) + 1]\, dx.
\end{equation}  
Here, $(h \ast \alpha)(x) = \int h(t)\alpha(x - t)\,dt$ is the convolution of $h$ with $\alpha$. Thus, any term in $N(\{f\})$ or $D(\{f\})$ involving $f$ contributes a term of this form. Summing across terms yields a linear functional of the form $  \int (h \ast \alpha)(x)\, \Phi(x)\, dx,
$ where $\Phi(x)$ involves expressions like $\log(f \ast \alpha)(x)$.


The rate functional takes the form $  \Upsilon(\ell) = \frac{N(\ell)}{D(\ell)},
$ and by the quotient rule, \cite{khandani2025} states that finding the best $f$ amounts to finding a stationary point of $\Upsilon(\ell)$:
\[
  \Upsilon'(\ell) = \frac{N'(\ell)\,D(\ell) - D'(\ell)\,N(\ell)}{(D(\ell))^2}.
\]
Taking the first-order (Gateaux) derivative with respect to an arbitrary perturbation $h$ corresponds to the Euler–Lagrange condition in variational calculus: at an optimum, the slope must vanish in every admissible direction—otherwise, a small step could further improve the objective~\cite{Nguyen_2024}. Thus, maximizing the functional reduces to requiring that the linear term in $\ell$ vanish for all valid $h$.

At $\ell=0$, set $N(0) = N_0$ and $D(0) = D_0$ as constants, and enforce $\Upsilon'(0) = 0$ for all $h$, yielding a condition of the form
\[
  \int (h \ast \alpha)(x)\, [\cdots]\, dx = 0
\]
for every admissible $h$. The solution must also include Lagrange multipliers to satisfy mean and power constraints.

Now, besides the stationarity of $\Upsilon(\ell)$, each codeword density $f$ must also satisfy $\int f(x)\,dx = 1,
\text{ } \int x^2\,f(x)\,dx \le P_{f}.$

Applying the variation $f \mapsto f + \ell h$ requires the following:
\begin{align}
  \int (f + \ell h) = 1
  \implies
  \int h(x)\,dx = 0,
\label{eq:mean}
\text{ and } \\
  \int x^2 (f + \ell h) = \int x^2 f(x)\,dx
 \implies
  \int x^2 h(x)\,dx = 0
  \label{eq:h_constraints}
\end{align}
If $f$ saturates the power constraint, then $\int x^2 f(x),dx = P_f$; otherwise, the associated Lagrange multiplier is zero. Theorem 6 in \cite{khandani2025} further shows that optimal codebook densities are zero-mean, and deviations from this lead to suboptimal power use. Therefore, we assume each codeword uses its full power and treat the constraint as an equality for brevity. Therefore, the \emph{net derivative} to be set to zero is:
\begin{equation}
  \Upsilon'(0) + \lambda \int h(x)\,dx
+ \mu \int x^2\ h(x)\,dx = 0
\label{eq:net_obj}
\end{equation}
where $\lambda$ and $\mu$ are Lagrange multipliers ensuring the constraints \eqref{eq:mean}, \eqref{eq:h_constraints} are satisfied for all admissible perturbation functions $h$. Since this equality must hold for all such $h$, the kernel multiplying $(h \ast \alpha)(x)$ must lie in the span of ${1, x^2}$, implying a specific form for $\ln(f\ast \alpha)$. Hence, it follows that:
\begin{equation}
  \ln\bigl((f \ast \alpha) (x)\bigr) \;=\; \alpha_1\,(-x^2) + \alpha_0
\label{eq:ln_quad}
\end{equation}
so every convolution $(f \ast \alpha)$ must be Gaussian, and by invertibility of the linear mixing (as discussed in \ref{subsec:conv}), each codebook density $f$ is Gaussian as well. A similar quadratic-kernel argument underlies the strong Entropy-Power Inequality in \cite{CourtadeStrongEPI2018}. Additionally, a standard result in functional analysis \cite{gelfand1963calculus} states that if $\ln(f \ast \alpha)$ deviates nontrivially from a quadratic polynomial, then a suitable perturbation $h$ will yield a nonzero derivative. Therefore, exponentiating yields:
\begin{equation}
 (f \ast \alpha)(x) = 
  C\exp(-\alpha_1^2/2),
\end{equation} 
which is a (zero-mean) \emph{Gaussian} (function) in $x$. Thus, the \emph{only} stationary solution to the derivative condition is that each relevant convolution term $(f\ast \alpha)$ is a Gaussian kernel.






\subsection{From Convolution to Individual Codewords}
\label{subsec:conv}

If $(f \ast \alpha)$ is Gaussian for multiple distinct $\alpha$ or for a set of sum combinations that collectively invert the codeword-to-sum mapping, then each codeword $f$ must have Gaussian marginals. The earlier argument implies that expressions like $V_1 + \sqrt{b} V_2$ (in the 2-user case) are Gaussian. Since these combinations can be collected into a random vector, and the optimal solution involves jointly Gaussian variables, it follows that $X_1, X_2$ are Gaussian as well.

For instance, consider successive decoding at $Y_2$ involving the composite random variables $C_1$ to $C_4$:
\begin{align}
C_1 &= \sqrt{a} U_1 + \sqrt{a} V_1 + U_2 + V_2 \\
C_2 &= \sqrt{a} U_1 + \sqrt{a} V_1 + V_2 \\
C_3 &= \sqrt{a} V_1 + V_2 \\
C_4 &= \sqrt{a} V_1
\end{align}
These form a linear system with coefficient matrix
$\begin{bmatrix}
\sqrt{a} & \sqrt{a} & 1 & 1 \\
\sqrt{a} & \sqrt{a} & 0 & 1 \\
0 & \sqrt{a} & 0 & 1 \\
0 & \sqrt{a} & 0 & 0
\end{bmatrix}$, which is invertible for all $a \neq 0$. Therefore, if the random vector $W = [C_1, C_2, C_3, C_4]^\top$ is jointly Gaussian, then so is $X = M^{-1}W$, implying that the underlying codeword components are jointly Gaussian. Hence, having shown that the sums in equations (11)-(14) are Gaussian, it follows that the vector $(U_1, V_1, U_2, V_2)$ is jointly Gaussian, or each sub-user component is \emph{itself} Gaussian. Repeating this reasoning for all user pairs leads to the conclusion that \emph{all codewords} must be Gaussian. This argument generalizes to the multi-user case via Entropy Power Inequality (EPI) methods~\cite{bhattacharya2025informationtheoreticefficientcapacityregion}. Next, in Section~IV, we present and delve into the details of our novel algorithm that solves Eq.~\eqref{eq:opt}.

\section{Proposed Algorithm}
\label{sec:methodology}


This section presents the core methodology for solving the \texttt{minPIC} problem. Although the high-level goal: minimizing total transmit power subject to rate constraints (Section~\ref{sec:prob-formulate})—is conceptually straightforward, the problem is structurally complex due to the interplay between (i) SIC decoding order at each receiver and (ii) power allocation across sub-user components. Our multi-stage approach: (I) applies Lagrangian duality to impose a convexity-compatible sub-stream order; (II) introduces auxiliary interference-bounding variables and log-det LMIs to relax SINR constraints; and (III) unifies the $U$ receivers’ capacity constraints via an intersection-of-unions framework, yielding a tractable convex (or partially convex) relaxation. We also address the challenges of jointly enforcing capacity constraints across all $U$ receivers.

\subsection{Decoding Order via Dual Variables}
\label{sec:method:dual}

The \texttt{minPIC} formulation in \eqref{eq:opt} lets each receiver pick a decoding order over all $U^{2}$ sub‑user components, yet \emph{convexity} of the IC capacity region~\cite{cioffi_gdfe_chapter} severely restricts which orders can be \emph{globally} optimal.  
From \eqref{eq:opt}, the Lagrangian can be written as follows:
\begin{align}
\label{eq:lagr}
\mathcal{L} &= \sum_{n=0}^{N-1} \sum_{u=1}^{U} w_u \, \text{tr} \left\{ \sum_{j=1}^{U} \mathbf{R}_{xx}(u,j,n) \right\} + \\
&\sum_{u=1}^{U} \lambda_u \left( b_{\min,u} - \sum_{j=1}^{U} b_{u,j} \right) &\nonumber
\end{align}
where the dual variables $\{\lambda_u\}$ measure the sensitivity to each user’s rate requirement.  
The optimal SIC decoding order at each receiver is uniquely specified by the dual variables $\{\lambda_u\}$ via complementary‐slackness conditions \cite{shi2015large}. In particular, at receiver $k$, larger values of $\lambda_u$ imply earlier decoding of user $u$’s sub‐streams. Thus, the $U$ numbers $\{\lambda_u\}$ alone impose a consistent partial order on all $U^{2}$ sub‑streams; any order that violates this ranking contradicts the \textit{convexity} and is sub‑optimal.

\begin{figure}
    \centering
    \includegraphics[width=\linewidth]{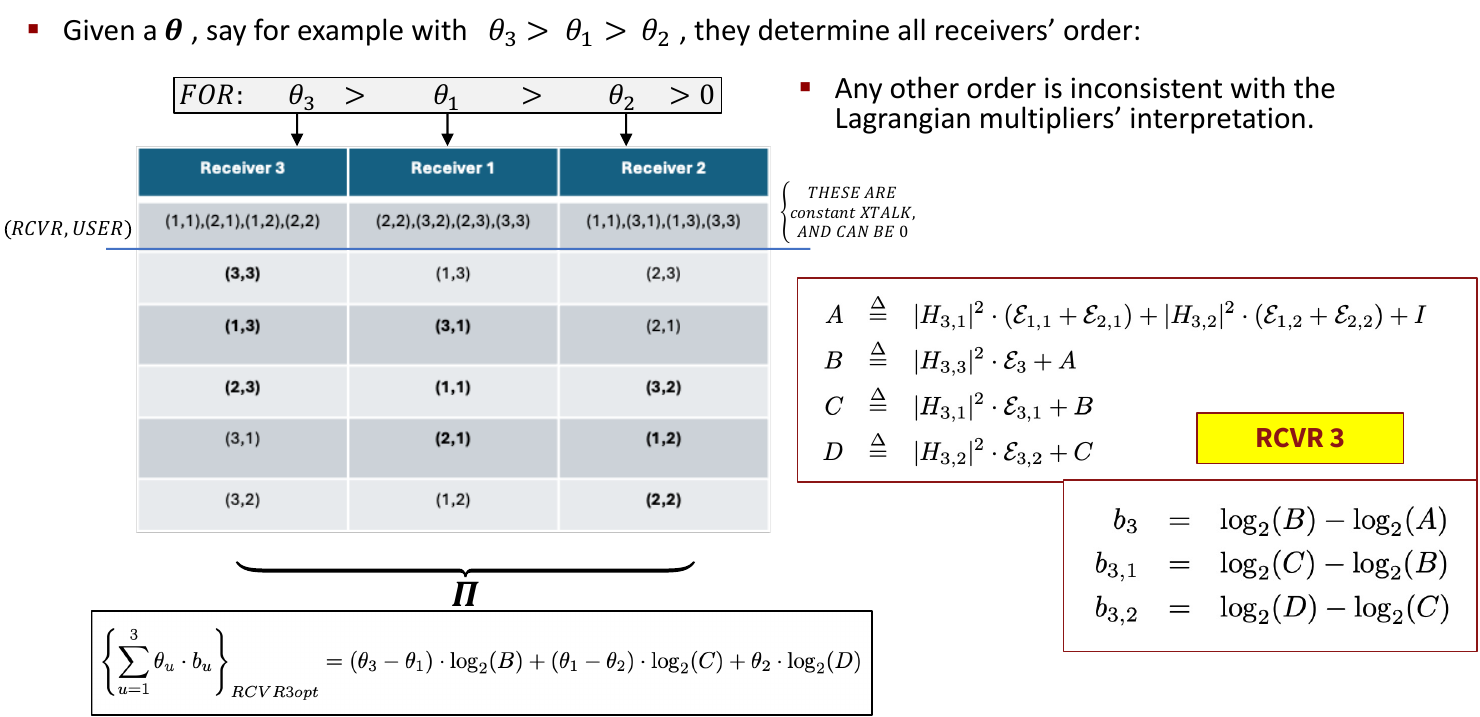}
    \caption{Example of Decoding Order being derived for 9 sub-user components, from a 3-dimensional $\theta$ vector}
    \label{fig:decodingOrder}
\end{figure}

\paragraph*{Illustrative example ($U=3$)}  
As illustrated in Fig. \ref{fig:decodingOrder}, if $\lambda_3>\lambda_1>\lambda_2>0$, then receiver $k$ first decodes all sub‐streams of user 3, then user 1, and finally user 2. Define:
\begin{align}
A &\triangleq |H_{3,1}|^2(\varepsilon_{1,1}+\varepsilon_{2,1}) + |H_{3,2}|^2(\varepsilon_{1,2}+\varepsilon_{2,2}) + I,\\
B &\triangleq |H_{3,3}|^2\cdot\varepsilon_3 + A,\text{ }
C \triangleq |H_{3,1}|^2\cdot\varepsilon_{3,1} + B, \\
D &\triangleq |H_{3,2}|^2\cdot\varepsilon_{3,2} + C
\end{align}

\vspace{-5pt}
and the incremental rates as:
\begin{align}
b_{3}=\log_{2}\frac{B}{A}, \text{ }
b_{3,1}=\log_{2}\frac{C}{B}, \text{ }
b_{3,2}=\log_{2}\frac{D}{C}
\end{align}
The resulting weighted sum‑rate is therefore 
$\Pi = \sum_{u=1}^3 \theta_u b_u
=(\theta_3 - \theta_1)\log_2 B+(\theta_1 - \theta_2)\log_2 C + \theta_2\log_2 D$.

Any other decoding order would \textbf{violate} $\{\lambda_u\}$‑induced ordering, therefore cannot be the global optimum.

\subsection{Nonconvexity of the Fixed-Order Formulation}
\label{sec:method:nonconvex}
Although the \texttt{minPIC} constraints under a fixed SIC order $\pi$ admit an explicit signal to interference and noise ratio (SINR)‐based form, the problem is generally nonconvex~\cite{7248899}. In particular, for sub-user $(k,j)$ its SINR after canceling earlier streams is 
\begin{equation}
\text{SINR}_{(k,j)}(\mathbf{p}) \;=\; \frac{p_{k,j}\,\Gamma_{k,j}}{\sigma^2 \;+\; \sum_{\ell \leq (k,j)} p_{\ell}\,\Gamma_{\ell}},
\label{eq:SINR_nonconvex}
\end{equation}
where $\Gamma_{k,j}$ is the channel‐gain norm. Since power terms appear in both num./denom. terms across multiple SINRs, the resulting problem is \textbf{nonconvex}.  Hence, for any fixed decoding order $\Pi$, the \texttt{minPIC} problem can be re-written as:
\begin{equation}
\label{eq:orderaware}
\begin{aligned}
\min_{\{\mathbf{R}_{xx}(i,j,n)\succeq0\}}\;&\sum_{n=0}^{N-1}\sum_{u=1}^U w_u\,\mathrm{tr}\Bigl\{\sum_{j=1}^U\mathbf{R}_{xx}(u,j,n)\Bigr\}\\
\text{s.t.}\quad
& \text{C1: }\sum_{j=1}^U\sum_{n=0}^{N-1}b_{u,j,n}\;\ge\;b_{\min,u},\text{ }\forall u,\\
&\text{C2: }\mathbf{R}_{xx}(u,j,n)\succeq0,\text{ }\forall\,u,j,n,
\end{aligned}
\end{equation}
where each sub‐stream’s rate on block~$n$ is
\[
b_{i,j,n}
=\log_{2}\!\Bigl(1 + \frac{\bigl|H_{r,i}\,\mathbf{R}_{xx}(i,j,n)\,H_{r,i}^{H}\bigr|}
{\sigma^{2} \;+\;\sum_{(p,q)}
\bigl|H_{r,p}\,\mathbf{R}_{xx}(p,q,n)\,H_{r,p}^{H}\bigr|}\Bigr) 
\]
Note that in the above expression we choose $(p,q): \Pi_{r}(p,q)\ge\Pi_{r}(i,j)$ (sub‐streams whose decoding index at receiver $r$ is $\ge$ $(i,j)$). Since each covariance matrix appears in both the numerator and denominator of the log‐SINR expressions (summed over all sub‐streams decoded at or after $(i,j)$), making the feasible set nonconvex. Prior methods~\cite{9109742,sun2018learning,shi2011wmmse} based on deep learning or alternating‐maximization heuristics often yield suboptimal or infeasible rate solutions under heavy congestion. In contrast, we pursue a convex relaxation that retains high fidelity even under stringent traffic demands.

\subsection{Decoupling Power Allocation from Decoding Order}
\label{sec:method:decoupling}

A natural question is whether \texttt{minPIC} admits a two‐stage solution: (i) allocating power $\mathbf{p}$ across subcarriers and (ii) selecting each receiver’s SIC order $\pi_i$. In the single‐receiver MAC, this succeeds since 
\begin{align*}
  \bigcup_{\pi}\bigl(\text{MAC region under }\pi\bigr)
  =
  \bigcup_{\pi}\bigcup_{\mathbf{p}}\{\mathbf{R}\},
\end{align*}
i.e. the overall capacity region is the convex hull of the regions achievable under each decoding order. Hence, power and order can be optimized independently, as in \textbf{minPMAC}~\cite{bhattacharya2024optimalpowerallocationtime}.

In the $U$‐receiver IC, the global capacity region is  
\[
  \bigcup_{\pi_1,\dots,\pi_U}
  \;\bigcap_{i=1}^U
  \bigl(\text{MAC region at receiver }i\text{ under order }\pi_i\bigr).
\]
Because one cannot time‐share decoding orders independently per receiver, the union over all order tuples and the intersection across receivers do not commute. Therefore, no “power‐then‐order” or “order‐then‐power” decomposition exists for the IC, and joint optimization of $\mathbf{p}$ and $\{\pi_i\}$ is necessary, making the problem much more challenging than \textbf{minPMAC}.

\subsection{Convex Relaxation via Log‐Det and Auxiliary Variables}
\label{sec:method:relaxation}

We now introduce auxiliary variables \(c_{r,n}\) per receiver \(r\) and block \(n\) to upper‐bound interference from undecoded sub‐streams. Under a fixed SIC order \(\Pi\), the original rate constraint for sub‐stream \((i,j)\) at receiver \(r\) on block \(n\),
\[
b_{i,j,n} + \sum_{(p,q)} b_{p,q,n}
\;\le\;
\log_{2}\!\bigl|\mathbf{I} 
+\mathbf{H}_{r,i}\,\mathbf{R}_{xx}(i,j,n)\,\mathbf{H}_{r,i}^{H}
+\mathbf{I}_{\text{unc},r,n}\bigr|,
\]
is relaxed by decoupling undecoded interference into \(c_{r,n}\):
\begin{align}
b_{i,j,n} + \sum_{(p,q)} b_{p,q,n} + c_{r,n}
&\le
\log_{2}\!\bigl|\mathbf{I} + \mathbf{H}_{r,i}\,\mathbf{R}_{xx}(i,j,n)\,\mathbf{H}_{r,i}^{H}\bigr|,
\label{eq:relaxed1}\\
c_{r,n} &\le
\log_{2}\!\bigl|\mathbf{I} + \mathbf{H}_{r,\mathrm{unc}}\,\mathbf{R}_{\mathrm{unc}}\,\mathbf{H}_{r,\mathrm{unc}}^{H}\bigr|.
\label{eq:relaxed2}
\end{align}
Here, \(\mathbf{H}_{r,\mathrm{unc}}\) and \(\mathbf{R}_{\mathrm{unc}}\) aggregate channels and covariances of all sub‐streams not decoded by \(r\). Using standard log‑det LMI representations~\cite{boyd1994linear}, these constraints yield a convex relaxation of the original SINR‐coupled formulation.
\[
\log\!\bigl|\mathbf{I} + \mathbf{H}\mathbf{R}_{xx}\mathbf{H}^{H}\bigr|
\;=\;\max_{\boldsymbol{\Theta}\succeq\mathbf{I}+\mathbf{H}\mathbf{R}_{xx}\mathbf{H}^{H}}
\;\log\!\bigl|\boldsymbol{\Theta}\bigr|,
\]
we convert \eqref{eq:relaxed1}–\eqref{eq:relaxed2} into an LMI-compatible formulation using \emph{log-det constraints} in $(\mathbf{R}_{xx},c,\boldsymbol{\Theta})$. The \textbf{final}, \textbf{reformulated} \textbf{optimization problem} can now be stated as:
\begin{equation}\label{eq:final_relaxed}
\begin{aligned}
\min_{\substack{R_{xx}(i,j,n)\succeq0\\c_{r,n}\ge0}}
\sum_{n=0}^{N-1}\sum_{u=1}^U\Bigl(w_u\,\mathrm{tr}\{\sum_{j=1}^U R_{xx}(u,j,n)\}-\lambda\,c_{u,n}\Bigr)\\
\text{s.t.}\quad
\text{C1: }\sum_{j=1}^U\sum_{n=0}^{N-1}b_{u,j,n}\;\ge\;b_{\min,u},\text{ C2: }R_{xx}(i,j,n)\succeq0, \\
\text{C3: }b_{i,j,n} + \sum_{(p,q)} b_{p,q,n} + c_{r,n}
\le \log_{2}\bigl|\mathbf{I} + H_{r,i}R_{xx}(.)H_{r,i}^H\bigr|,\\
\text{C4: }c_{r,n}\le \log_{2}\bigl|\mathbf{I}+H_{r,\mathrm{unc}}\,R_{\mathrm{unc}}\,H_{r,\mathrm{unc}}^H\bigr|\ \forall\,r,u, i, j, n.
\end{aligned}
\end{equation}

Since each \(\log\det\) term is jointly concave in \(\mathbf{R}_{xx}\) and linearized via \(\boldsymbol{\Theta}\succeq\cdots\), the relaxation is \textbf{convex}. Empirically, at optimality $c_{r,n}=\log\bigl|\mathbf{I}+\mathbf{H}_{r,\text{unc}}\mathbf{R}_{\text{unc}}\mathbf{H}_{r,\text{unc}}^{H}\bigr|$, 
so the bound is often tight even under heavy interference. This reformulation converts ratio‐based SINR constraints into differences of log terms, eliminating fractional terms and yielding a tractable convex program. We also observe \textbf{strong duality} numerically, ensuring reliable convergence; without it, Slater’s condition would require extra constraints for dual feasibility~\cite{boyd2004convex}.

\subsection{Choosing Weight Parameters and Binary Search for Tight Constraints}
\label{sec:method:lambda}

The convex relaxation’s Lagrangian takes the form: 
\begin{equation}
\min_{\mathbf{p},\,\{c_{k,j}\}} P_{\text{total}}(\mathbf{p}) \;+\; \lambda \sum_{k,j}\bigl(c_{k,j} - R_{k,j}^{\min}\bigr),
\label{eq:relaxed_lagrangian}
\end{equation}
and balances power efficiency against strict rate satisfaction:  $\lambda$ being too large wastes power, and it being too small allows user-rate violations. To find the minimal $\lambda$ that enforces all $c_{k,j}\ge R_{k,j}^{\min}$ tightly, we use a simple bisection:

\begin{enumerate}
  \item Set $\lambda_{\text{low}}=0$ and a large $\lambda_{\text{high}}$ ensuring feasibility.
  \item Repeat until $\lambda_{\text{high}}-\lambda_{\text{low}}<\epsilon$:
  \begin{itemize}
    \item Let $\lambda=(\lambda_{\text{low}}+\lambda_{\text{high}})/2$ and solve \eqref{eq:relaxed_lagrangian}.
    \item If all rate constraints hold with slack, set $\lambda_{\text{high}}=\lambda$, else $\lambda_{\text{low}}=\lambda$.
  \end{itemize}
\end{enumerate}

In our implementation, once the final $\lambda$ is found so that $c_{k,j} \approx R_{k,j}^{\min}$ for each sub-user component, the solution has an accurate power allocation that respects the partial MAC constraints. This bracketing ensures that the result does not overshoot or undershoot the required rates.

\vspace{-4.1pt}

\subsection{Reducing Sub-User Degrees of Freedom}
\label{sec:method:reduction}
A final (yet crucial) refinement involves reducing the dimensionality of the sub-user power allocation space. Structural dependencies in the SIC chain rule imply that, although \texttt{minPIC} defines $U^2$ power variables $p_{k,j}$, only $U^2 - U + 1$ degrees of freedom (or fewer in MIMO) are needed. Aggregating redundant sub-streams into ``private” or ``common” groups preserves optimality while cutting the primal dimension.  The final algorithm proceeds as follows:
\begin{enumerate}
  \item Collapse power variables from $U^2$ to $U^2 - U + 1$ (or the equivalent MIMO count).
  \item For a fixed SIC order, build \& solve the convex relaxation with log‐det LMIs, auxiliary variables $\{c_{k,j}\}$ (as in \eqref{eq:final_relaxed}), and binary search over $\lambda$ (\S\ref{sec:method:lambda}).
  \item Optimize decoding orders: either by full enumeration or via the dual‐variable‐induced ranking. (\S\ref{sec:method:dual}).
\end{enumerate}

\section{Numerical Evaluation}
\label{sec:performance}

\subsection{\texttt{minPIC} demo}
In both the examples below, we assume that minimum rate requirements are $R_k^{\min}=0.5$ bits/dimension.
\subsubsection*{2-User Case} The channel matrix is:
\[
H \;=\;\begin{bmatrix}
0.4 & 0\\
0.9 & 1
\end{bmatrix}.
\]
The bisection on the penalty weight converged to $\lambda \approx 16.80$. The resulting optimal power allocations are shown in Table~\ref{tab:example_2user_power} while the achieved data rates are $[0.500, 0.500]$ (in bits/dim) for both users.

\vspace{-10pt}

\begin{table}[!ht]
\centering
\caption{Optimal Sub-User Power Allocations for $U=2$}
\label{tab:example_2user_power}
\begin{tabular}{c|cc}
\hline
\textbf{User $k$ / Sub-user $j$} & \textbf{1} & \textbf{2} \\
\hline
1 & 11.52 & 1.17 \\
2 &  8.74 & 1.59 \\
\hline
\end{tabular}
\end{table}



\subsubsection*{3-User Case} The channel matrix is:
\[
H \;=\;
\begin{bmatrix}
1      & 0.9    & 0.001\\
0.001  & 1      & 0.001\\
0.001  & 0.001  & 1
\end{bmatrix}.
\]
The bisection on the penalty weight converged to $\lambda \approx 5.016$.

The resulting optimal power allocations are shown in Table~\ref{tab:example_3user} while the achieved data rates are $[0.500, 0.500, 0.500]$ (in bits/dim) for all users.


\begin{table}[!ht]
\centering
\caption{Optimal Sub-User Power Allocations for $U=3$}
\label{tab:example_3user}
\begin{tabular}{|c|ccc|}
\hline
\textbf{User $k$ / Sub-user $j$} & \textbf{1} & \textbf{2} & \textbf{3} \\
\hline
1 & 2.58 & 0.00 & 0.00 \\
2 & 0.00 & 1.95 & 0.00 \\
3 & 0.02 & 0.00 & 0.98 \\
\hline
\end{tabular}
\end{table}





Overall, these experiments confirm the viability of the proposed convex relaxation plus multi-order approach in minPIC, consistently achieving the targeted user rates while minimizing total transmit power. \textbf{To the best of our knowledge, we are the first to provide optimal design for a 3-user IC.}


\subsection{Extended single‑subcarrier study:}
To stress‑test \emph{minPIC} across the full interference spectrum, we evaluate the five representative single‑tone scenarios listed in Table\ref{tab:test_suite}.
They cover (i) an \emph{almost‑orthogonal} link where \emph{minPIC} collapses to water‑filling (Case~A), (ii) \emph{moderate symmetric} coupling (B), (iii) \emph{highly‑asymmetric strong} interference (C), and two $3$‑user topologies—(iv) a \emph{weakly‑coupled} network (D) and
(v) a \emph{one‑dominant‑link} setting (E). All users require the same QoS target $R_{\min}=0.5$ bit/Hz, and $\sigma^{2}=1$. Table \ref{tab:test_suite} reports the total transmit power produced by \emph{minPIC}; analytic OMA estimates (not shown for
brevity) lie 20–40\,\% higher for Cases B–E. The results confirm that \emph{minPIC} seamlessly reduces to water‑filling when interference is negligible (A), yields the largest savings—over 40\,\% versus OMA—when a single dominant cross‑link is present (E).

\begin{table}[t]
  \centering
  \caption{Focused single‑subcarrier test‑suite
           ($\sigma^{2}=1,\;R_{\min}=0.5$ bit/Hz/user)}
  \label{tab:test_suite}
  \renewcommand{\arraystretch}{1.1}
  \setlength{\tabcolsep}{4pt}
  \begin{tabular}{|c|c|c|c|c|}
    \hline
    \textbf{ID} & \textbf{$U$} &
    \textbf{Non‑zero $H_{ij}$ entries} &
    \textbf{Regime} &
    $\sum P_{\text{minPIC}}$ (W) \\
    \hline
    A & 2 & $H_{01}=H_{10}=10^{-6}$    & almost‑orthogonal & 1.76 \\
    B & 2 & $H_{01}=H_{10}=10^{-2}$    & moderate sym.     & 2.00 \\
    C & 2 & $H_{01}=0.9,\;H_{10}=10^{-5}$ & strong asym.   & 4.33 \\
    D & 3 & all off‑diag.\ $10^{-5}$    & 3‑user weak       & 2.43 \\
    E & 3 & $H_{01}=0.9$, others $10^{-3}$ & 3‑user dom.\ link & 4.55 \\
    \hline
  \end{tabular}
\end{table}

\section{Limitations and Conclusion}
\label{sec:concld}
Existing OMA, NOMA and RSMA schemes \emph{fix} the SIC order and optimise power only, or restrict the search to at most two users. \emph{This paper is the first to \textit{jointly and optimally} design both the SIC decoding order \textbf{and} the full $U^{2}$ power allocation for an \textit{arbitrary} $U$‑user interference channel,} thereby attempting the \emph{first algorithmic realisation of the exact IC capacity boundary.}
We also note that the present study is quite limited to (i) a single subcarrier,
(ii) small user counts, and (iii) flat‑fading links with perfect
CSI. Addressing these questions will turn
minPIC from a theoretically optimal single‑tone engine into a
deployable 6G interference‑management solution.





\bibliographystyle{ieeetr}
\bibliography{references}

\end{document}